\documentclass[10pt,a4paper]{article}
\usepackage[T2A]{fontenc}
\usepackage[utf8]{inputenc}
\usepackage[english]{babel}
\usepackage{amssymb,graphicx}
\usepackage{amsmath,amsfonts}
\usepackage{amsthm}
\usepackage{framed}
\usepackage{xcolor}
\usepackage{fullpage}
\usepackage[makeroom]{cancel}
\usepackage[export]{adjustbox} 
\usepackage{microtype}
\usepackage{hyperref}

\title{\textbf{Hanany-Witten brane crossing\\
    and Ding-Iohara-Miki algebra}}
\author{Yegor Zenkevich\thanks{yegor.zenkevich@gmail.com}\\
  {\small\textit{SISSA, via
      Bonomea 265, 34136 Trieste, Italy,}}\\
  {\small\textit{INFN, Sezione di Trieste,}}\\
  {\small\textit{IGAP, via Beirut 2/1, 34151 Trieste, Italy,}}\\
  {\small\textit{ITEP, Bolshaya Cheremushkinskaya street 25, 117218
      Moscow, Russia,\footnote{On leave.}
    }}\\
  {\small\textit{ITMP MSU, Leninskie gory 1, 119991 Moscow,
      Russia,\footnotemark[\value{footnote}] }}} \date{}
\begin{document}
\maketitle
\vspace{-46ex}
\noindent
{\textit{To T.} \hfill ITEP/TH-29/22}
\vspace{40ex}

\begin{abstract}
  We further develop the correspondence between representations of
  Ding-Iohara-Miki (DIM) algebra and Type IIB branes. In particular we
  explicitly compute the Hanany-Witten type 5-brane crossing operator
  which plays the role of the $R$-matrix and has interesting
  combinatorial properties. We explore the corresponding lattice
  integrable models and notice a possible connection with statistics
  of plane partitions.
\end{abstract}

\section{Introduction}
\label{sec:introduction}

In~\cite{Zenkevich:2018fzl} we have introduced a correspondence
between branes of Type IIB string theory and representations of the
Ding-Iohara-Miki (DIM) algebra~\cite{DIM}. The long known example of
this correspondence is the refined topological
vertex~\cite{Awata:2005fa}, which can be viewed as a triple junction
of 5-branes or as an intertwining operator of DIM
representations~\cite{AFS}. In~\cite{Zenkevich:2020ufs} we have
suggested another type of brane interaction --- the crossing in which
the branes don't join, but instead pass behind each other in different
planes. The algebraic meaning of the crossing is an $R$-matrix of the
DIM algebra taken in the representations corresponding to the two
crossing branes. The slogan of the Type IIB/DIM correspondence
therefore can be formulated as
\begin{gather}
  \label{eq:2}
  \text{Branes} = \text{DIM representations,}\\
  \text{Brane junctions} = \text{DIM intertwining operators,}\label{eq:9}\\
  \text{Brane crossings} = \text{DIM $R$-matrices.}\label{eq:10}
\end{gather}

In this sense, the brane setup resembles the $3d$ (or $4d$)
Chern-Simons theory in which Wilson lines cross in the $2d$ projection
of the theory giving rise to $R$-matrices of the quantum group or
quantum affine Lie algebra respectively~\cite{Costello}.

Naturally, the Type IIB case in much more complicated that the
Chern-Simons field theory. In particular, there are many types of
branes and therefore many different types of crossings. In the current
paper we focus on the crossings between 5-branes. Even this class,
however, is very rich. To analyze possible crossings let us look at
the setup Type IIB setup summarized in table below.
\begin{equation}
  \label{eq:1}
    \begin{array}{l|l|ccc|cc|c|c}
      &&&&&\multicolumn{2}{c|}{\text{picture}}&\tau&\\
      \text{Brane}& \text{DIM rep} &\textcolor{blue}{\mathbb{C}_q}&\textcolor{red}{\mathbb{C}_{t^{-1}}}&\textcolor{violet}{\mathbb{C}_{t/q}}&\mathbb{R}_x&\mathbb{R}_y&\mathbb{R}_{\tau}&S^1\\
    \hline
    \textcolor{violet}{\mathrm{D5}_{q,t^{-1}}} &
    \textcolor{violet}{\mathcal{F}^{(0,1)}_{q,t^{-1}}}&--&--&&&-&&-\\
    \textcolor{red}{\mathrm{D5}_{q,t/q}} &
    \textcolor{red}{\mathcal{F}^{(0,1)}_{q,t/q}}&--&&--&&-&&-\\
    \textcolor{blue}{\mathrm{D5}_{t^{-1},t/q}} &
    \textcolor{blue}{\mathcal{F}^{(0,1)}_{t^{-1},t/q}}&&--&--&&-&&-\\
    \hline
    \textcolor{violet}{\mathrm{NS5}_{q,t^{-1}}} &
    \textcolor{violet}{\mathcal{F}^{(1,0)}_{q,t^{-1}}}&--&--&&-&&&-\\
    \textcolor{red}{\mathrm{NS5}_{q,t/q}} &
    \textcolor{red}{\mathcal{F}^{(1,0)}_{q,t/q}}&--&&--&-&&&-\\
    \textcolor{blue}{\mathrm{NS5}_{t^{-1},t/q}} &
    \textcolor{blue}{\mathcal{F}^{(1,0)}_{t^{-1},t/q}}&&--&--&-&&&-
  \end{array} 
\end{equation}
The 5-branes carry $(p,q)$ charges, and their slope in the plane of
the picture is tied with their charge because of supersymmetry
constraints. The $(1,0)$ 5-brane is the D5 brane, while $(0,1)$ is
the NS5 brane. In Table~\eqref{eq:1} the indices $q$, $t^{-1}$ and
$t/q$ carried by the $\mathbb{C}$-planes signify the $6d$ equivariant
Omega-background. The colors indicate are used to the same effect. The
5-branes span pairs of complex planes, carry the corresponding indices
and their color indicates the one out of three planes that they
\emph{don't} span.

According to the correspondence~\cite{Zenkevich:2018fzl} branes
correspond to representations of the DIM algebra. In particular
5-branes correspond to Fock representations, which conveniently happen
to carry the labels corresponding to the $(p,q)$ charge (the slope of
the brane and of the representation) and the color corresponding
to the choice of two out of three equivariant parameters $q$, $t^{-1}$
and $t/q$.

Branes of the same color living at the same value of the $\tau$
coordinate can form triple junctions which correspond to DIM
intertwining operators, i.e.\ homomorphism from tensor products of a
pair of Fock representations to a single Fock representation or vice
versa. The network of such intertwiners correspond to a 5-brane web,
and the vacuum matrix element of the network reproduces the refined
topological partition function.

It is evident from the table that a D5 and an NS5 brane can cross in
projection on the plane of the picture $\mathbb{R}^2_{xy}$, while
having different $\tau$ coordinates. Algebraically the crossings
correspond to the DIM $R$-matrix exchanging the Fock
representations. There are two essentially different cases: either the
D5 and NS5 branes are of the same color or their colors are
different. Since the algebra is symmetric under permutation of
$(q,t^{-1},t/q)$ without loss of generality we can consider only two
types of D5-NS5 crossings: D5$_{q,t^{-1}}$--NS5$_{q,t^{-1}}$ and
D5$_{q,t^{-1}}$--NS5$_{q,t/q}$. In~\cite{Zenkevich:2020ufs} we have
considered a particular case of crossings with vacuum states on the
NS5 brane. In the present paper we remedy this situation and obtain
the crossing operators for \emph{general} external states. This will
allow us to combine the crossings into a lattice integrable model.

The D5$_{q,t^{-1}}$--NS5$_{q,t^{-1}}$ crossing can be obtained by
degenerating a pair of triple junctions~\cite{Zenkevich:2020ufs} which
correspond to refined topological vertices. We call this crossing the
\emph{degenerate resolution} and recall the resolution procedure in
sec.~\ref{sec:warm-up:-degenerate}.

The D5$_{q,t^{-1}}$--NS5$_{q,t/q}$ crossing cannot be ``resolved'' in
this way since the branes of different colors cannot form a triple
junction. One can notice that the way D5$_{q,t^{-1}}$ and
NS5$_{q,t/q}$ branes are placed in the setup~\eqref{eq:1} is the same
as in the classic works of Hanany and
Witten~\cite{Hanany:1996ie}. Indeed, the D5$_{q,t^{-1}}$ and
NS5$_{q,t/q}$ branes share only three spatial dimensions. We therefore
call this type of crossing the \emph{Hanany-Witten crossing.}
In~\cite{Zenkevich:2020ufs} we have shown that this type of crossing
indeed manifests the Hanany-Witten brane creation effect. To find the
explicit expression for the corresponding ``Hanany-Witen $R$-matrix''
for general external states one needs to resort to first principles
and check the intertwining property. We do this in
sec.~\ref{sec:hanany-witten-5}.

It turns out that the Hanany-Witten $R$-matrix has curious
combinatorial properties with many of its matrix elements
vanishing. In sec.~\ref{sec:comp-plane-part} we use this combinatorics
to investigate a lattice statistical model built using the
$R$-matrices and find that they are related to the plane
partitions. We present our conclusions in sec.~\ref{sec:conclusions}.

\section{Warm-up: degenerate resolution}
\label{sec:warm-up:-degenerate}
Let us start by briefly recalling the construction of the
D5$_{q,t^{-1}}$--NS5$_{q,t^{-1}}$ crossing
from~\cite{Zenkevich:2020ufs}. We start with the network of
intertwiners consisting of two triple junctions of 5-branes:
\begin{equation}
  \label{eq:3}
  C(u,w,Q) \quad =\quad \includegraphics[valign=c]{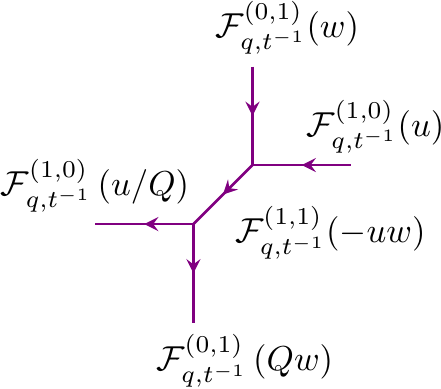}
\end{equation}
In the refined topological string this network would depict an open
amplitude on a resolved conifold geometry. We consider slightly more
general situation than in~\cite{Zenkevich:2020ufs} with nontrivial
states on the vertical external legs of the diagram.

We are not going to repeat the arguments of~\cite{Zenkevich:2020ufs}
leading to this, but there are two special values of $Q = \left(
  \frac{t}{q} \right)^{\pm 1/2}$ for which the operator $C(u,w,Q)$
degenerates into the DIM $R$-matrix acting in the corresponding Fock
representations:
\begin{equation}
  \label{eq:4}
  C \left(u,w, (t/q)^{\pm \frac{1}{2}}
  \right) = \mathcal{R}^{\pm 1}|_{\mathcal{F}^{(0,1)}_{q,t^{-1}}(w) \otimes
    \mathcal{F}^{(0,1)}_{q,t^{-1}}(u)}.
\end{equation}
The picture corresponding to the $R$-matrix is that of two 5-branes
crossing in the projection on $\mathbb{R}^2_{xy}$, but living at
different values of $\tau$:
\begin{equation}
  \label{eq:6}
  \mathcal{R}^{\lambda}_{\mu}(w,u) =  \left( \ldots \otimes \left\langle \mu, \sqrt{\frac{t}{q}} w \right| \right) \mathcal{R}|_{\mathcal{F}^{(0,1)}_{q,t^{-1}}(w) \otimes
    \mathcal{F}^{(0,1)}_{q,t^{-1}}(u)} \left( | \lambda, w\rangle
    \otimes \ldots \right) \quad = \quad \includegraphics[valign=c]{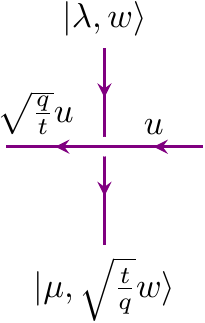}
\end{equation}
where $\lambda$ and $\mu$ are Young diagrams labelling the states in
the vertical Fock representation. As we have announced in the
sec.~\ref{sec:introduction} we call this type of crossing the
\emph{degenerate resolution,} since it is obtained by degenerating the
toric diagram of resolved conifold singularity.

Because degenerate resolution can be obtained by interpolation from
the non-degenerate resolution~\eqref{eq:3} its explicit form can be
written down explicitly. We just have to set $Q = \sqrt{\frac{t}{q}}$
in Eq.~\eqref{eq:3} and plug in the refined vertices
from~\cite{AFS}. The end result is:
  \begin{framed}
\begin{multline}
  \label{eq:5}
    \mathcal{R}^{\lambda}_{\mu}(w,u) =  N_{\lambda \mu}\left( \frac{q}{t} \right) u^{|\lambda|}
  \left( \frac{q}{u} \sqrt{\frac{t}{q}} \right)^{|\mu|}
  \frac{q^{n(\lambda^{\mathrm{T}})+ n(\mu^{\mathrm{T}})}}{f_{\lambda}
    C_{\lambda} C_{\mu}} \left( \frac{q}{t} \right)^{\frac{\hat{Q}}{2}} : \exp \left[ - \sum_{n \geq 1} \frac{w^n}{n} \frac{\left( 1-
        \left( t/q \right)^n\right)}{(1-q^n)} a_{-n} \right]\times\\
  \times \exp
  \left[ - \sum_{n \neq 0} \frac{(1-t^n) }{n} \left(
      \mathrm{Ch}_{\lambda}(q^{-n},t^n) - \left( \frac{t}{q}
      \right)^{\frac{|n|-n}{2}} \mathrm{Ch}_{\mu}(q^{-n},t^n)
    \right)w^{-n} a_n \right] :,
\end{multline}
\end{framed}
where bosonic modes $a_n$ satisfying
\begin{equation}
  \label{eq:8}
  [a_n, a_m] = n \frac{1-q^{|n|}}{1-t^{|n|}} \delta_{n+m,0}
\end{equation}
are understood to act on the horizontal Fock representations,
$(q/t)^{\hat{Q}/2}$ is the zero mode operator shifting the spectral
parameter $u$ of the horizontal representation by $\sqrt{q/t}$, the
symbol $:\ldots :$ denotes the normal ordering, and
\begin{gather}
  \label{eq:7}
  N_{\lambda \mu}^{(q,t^{-1})}(x) = \prod_{(i,j) \in \lambda} \left(  1 - x
    q^{\lambda_i - j} t^{\mu^{\mathrm{T}}_j - i + 1} \right) \prod_{(k,l) \in \mu} \left(  1 - x
    q^{-\mu_i + j-1} t^{-\lambda^{\mathrm{T}}_j + i} \right),\\
    n(\lambda^{\mathrm{T}}) = \sum_{(i,j)\in \lambda} (j-1),\qquad
    \qquad   f_{\lambda}(q,t^{-1}) = \prod_{(i,j)\in \lambda} \left( - q^{j-\frac{1}{2}}
    t^{\frac{1}{2} - i} \right),\\
  C_{\lambda}(q,t^{-1}) = \prod_{(i,j)\in \lambda} \left( 1 - q^{\lambda_i - j}
    t^{\lambda_j^{\mathrm{T}} - i +1} \right),\\
  \mathrm{Ch}_{\lambda}(q,t^{-1}) = \sum_{(i,j)\in \lambda} q^{j-1}\label{eq:33}
  t^{1-i}.
\end{gather}
The crossing~\eqref{eq:5} satisfies the intertwining property with DIM
algebra elements by construction, as in should according to the
dictionary~\eqref{eq:2}--\eqref{eq:10}. One also can check that a pair
of degenerate resolutions~\eqref{eq:5} satisfies Yang-Baxter or
$RTT$-relations with an extra DIM $R$-matrix acting on the vertical
legs~\cite{Awata:2016mxc}.

The crucial part of the degenerate resolution~\eqref{eq:5} is the
Nekrasov factor $N_{\lambda \mu} \left( \frac{q}{t} \right)$, which
satisfies certain selection rules, namely
\begin{equation}
  \label{eq:11}
  N_{\lambda \mu} \left( \frac{q}{t} \right) = 0\quad \text{unless} \quad \mu \subseteq \lambda.
\end{equation}
Therefore, if the ``boundary
condition'' $\lambda$ is fixed there is only a finite number of Young
diagrams $\mu$ which contribute to any amplitude involving the
crossing. We will exploit this property when we combine several
crossings into a statistical model in sec.~\ref{sec:comp-plane-part}

\section{The Hanany-Witten 5-brane crossing}
\label{sec:hanany-witten-5}
In this section we will determine the explicit form of the crossing
operator of a D5$_{q,t/q}$ and NS5$_{q,t^{-1}}$ branes, or the
Hanany-Witten crossing for arbitrary external states.

The crossing is represented by the following picture:
\begin{equation}
  \label{eq:12}
  \includegraphics[valign=c]{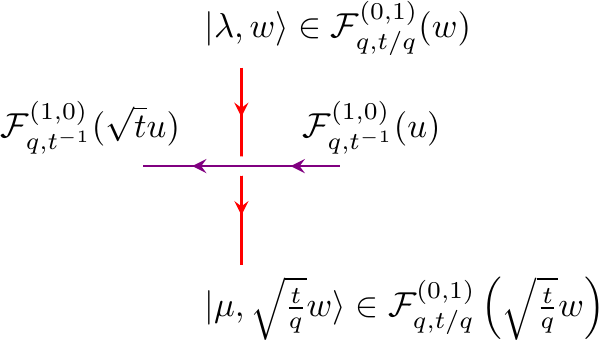} \hspace{-3em} = \quad
      \left( \ldots \otimes \left\langle \mu, \sqrt{\frac{t}{q}} w \right| \right) \mathcal{R}|_{\mathcal{F}^{(0,1)}_{q,t/q}(w) \otimes
    \mathcal{F}^{(0,1)}_{q,t^{-1}}(u)} \left( | \lambda, w\rangle
    \otimes \ldots \right) = \widetilde{\mathcal{R}}^{\lambda}_{\mu}(w,u)
\end{equation}
The vertical and horizontal branes are of different types, so there is
no way they can join together into a picture resembling
Eq.~\eqref{eq:3}. One needs to make a guess of the form of the
$R$-matrix and then check that it satisfies the intertwining relations
with the elements of the DIM algebra:
  \begin{equation}
  \label{eq:15}
\boxed{\mathcal{R}|_{\mathcal{F}^{(0,1)}_{q,t/q}(w) \otimes
    \mathcal{F}^{(0,1)}_{q,t^{-1}}(u)}\Delta(g)|_{\mathcal{F}^{(0,1)}_{q,t/q}(w) \otimes
    \mathcal{F}^{(0,1)}_{q,t^{-1}}(u)} = \Delta(g)|_{
    \mathcal{F}^{(0,1)}_{q,t^{-1}}(u) \otimes \mathcal{F}^{(0,1)}_{q,t/q}(w)} \mathcal{R}|_{\mathcal{F}^{(0,1)}_{q,t/q}(w) \otimes
    \mathcal{F}^{(0,1)}_{q,t^{-1}}(u)},}
\end{equation}
where $\Delta$ is the coproduct which we will write down explicitly
below and $g$ is any element of the DIM algebra. Notice that the order
of the representations on which the coproduct acts in the r.h.s.\ of
Eq.~\eqref{eq:15} is opposite to that in the l.h.s.

We assume that $\widetilde{\mathcal{R}}^{\lambda}_{\mu}(w,u)$ has the
following vertex operator form
\begin{equation}
  \label{eq:13}
  \widetilde{\mathcal{R}}^{\lambda}_{\mu}(w,u) = f^{\lambda}_{\mu}(w,u) \mathbf{R}^{\lambda}_{\mu}(w) =
  f^{\lambda}_{\mu}(w,u) t^{\frac{\hat{Q}}{2}} :\exp \left[ \sum_{n \neq 0}
    c_n w^{-n} a_n \right]:,
\end{equation}
where $f^{\lambda}_w(w,u)$ is a $\mathbb{C}$-valued function of the
spectral parameters, $c_n$ are some complex coefficients, $a_n$ are
the same modes as in Eq.~\eqref{eq:8} and $t^{Q/2}$ is the bosonic
zero mode shifting the horizontal spectral parameter $u$ by
$\sqrt{t}$. The change in the spectral parameter can be read off
directly from the universal $R$-matrix of the DIM algebra. This has
been done in~\cite{Zenkevich:2020ufs}, where
$\mathcal{\widetilde{R}}^{\varnothing}_{\varnothing}(w,u)$ has also
been found.

\subsection{The vertex operator part}
\label{sec:find-vert-oper}
To find $c_n$ let us consider the intertwining relation with the
generating currents $\psi^{\pm}(z)$ of DIM algebra. We will not write
DIM commutation relations which can be found for example in
Appendix A of~\cite{Zenkevich:2018fzl}. The coproduct acts on
$\psi^{\pm}(z)$ in the following way:
\begin{equation}
  \label{eq:14}
    \Delta(\psi^{\pm}(z)) = \psi^{\pm}\left(
    \gamma_{(2)}^{\pm \frac{1}{2}}z \right) \otimes \psi^{\pm}\left(
    \gamma_{(1)}^{\mp \frac{1}{2}}z \right),
\end{equation}
where $\gamma_{(1)} = \gamma \otimes 1$, $\gamma_{(2)} = 1 \otimes
\gamma$ and $\gamma$ is the ``horizontal'' central charge, which is
equal to $\sqrt{\frac{t}{q}}$ for $\mathcal{F}^{(1,0)}_{q,t^{-1}}$ and
$1$ for $\mathcal{F}^{(0,1)}_{q, t/q}$. Plugging the
coproduct~\eqref{eq:14} for $\psi^{-}(z)$ into Eq.~\eqref{eq:15} and
sandwiching between the $\langle\mu|$ and $|\lambda\rangle$ states we
get
\begin{multline}
  \label{eq:16}
  \mathbf{R}^{\lambda}_{\mu}(w,u)
  \psi^{-}(z)|_{\mathcal{F}^{(1,0)}_{q,t^{-1}}(u)} \frac{1}{\sqrt{t}}
  \exp \left[ \sum_{n \geq 1} \frac{1}{n} \left( \frac{t}{q}
    \right)^{-\frac{n}{4}} \left( \frac{w}{z} \right)^{-n} \left( 1 -
      t^n + \kappa_n \mathrm{Ch}_{\lambda} \left( q^{-n}, \left( t/q \right)^{-n} \right) \right) \right] \stackrel{?}{=}\\
  \stackrel{?}{=} \frac{1}{\sqrt{t}} \exp \left[ \sum_{n \geq 1}
    \frac{1}{n} \left( \frac{t}{q} \right)^{-\frac{n}{4}} \left(
      \frac{w}{z} \right)^{-n} \left( 1 - t^n + \kappa_n
      \mathrm{Ch}_{\mu} \left( q^{-n}, \left( t/q \right)^{-n} \right) \right) \right]
  \psi^{-}(z)|_{\mathcal{F}^{(1,0)}_{q,t^{-1}}(\sqrt{t} u)}
  \mathbf{R}^{\lambda}_{\mu}(w,u),
\end{multline}
where $\kappa_n = (1-q^n)(1-t^{-n})(1-(t/q)^n)$. To get
Eq.~\eqref{eq:16} we have used the action of DIM currents on the
representation $\mathcal{F}^{(0,1)}_{q,t/q}(w)$ (see
Appendix~\ref{sec:vert-repr-dim}). We have to notice that
Eq.~\eqref{eq:16} is only valid for $\lambda$ and $\mu$ such that
$f^{\lambda}_{\nu}(w,u) \neq 0$. As we will see in
sec.~\ref{sec:finding-prefactor} there are many pairs for which the
prefactor $f^{\lambda}_{\mu}(w,u)$ indeed vanishes, however this would not
affect the results of this section.

Next we use the expression for the current $\psi^{-}(z)$ in the
representation $\mathcal{F}^{(1,0)}_{q,t^{-1}}(u)$ (can be found
in~\cite{Zenkevich:2018fzl}), normal order the l.h.s. of
Eq.~\eqref{eq:16} and observe that the r.h.s.\ is already normal
ordered. We have
\begin{equation}
  \label{eq:18}
  \mathbf{R}^{\lambda}_{\mu}(w,u)
  \psi^{-}(z)|_{\mathcal{F}^{(1,0)}_{q,t^{-1}}(u)} = \exp \left[
    \sum_{n \geq 1} c_n \left( \frac{z}{w} \right)^n
    \frac{\kappa_n}{1-t^n} \left( \frac{q}{t} \right)^{\frac{n}{4}}  \right] :\mathbf{R}^{\lambda}_{\mu}(w,u)
  \psi^{-}(z)|_{\mathcal{F}^{(1,0)}_{q,t^{-1}}(u)}:
\end{equation}
Plugging Eq.~\eqref{eq:18} into Eq.~\eqref{eq:16} we finally get the
expression for $c_n$:
\begin{equation}
  \label{eq:17}
  \boxed{c_n = \frac{1-t^n}{n} \left[ \mathrm{Ch}_{\mu} \left( q^{-n},
      \left( t/q \right)^{-n} \right) - \mathrm{Ch}_{\lambda} \left(
      q^{-n}, \left( t/q \right)^{-n} \right) \right], \qquad
  \text{for }  n \geq 1.}
\end{equation}
Making a similar exercise for the current $\psi^{+}(z)$ we get the
remaining half of the coefficients $c_{-n}$:
\begin{equation}
  \label{eq:19}
 \boxed{ c_{-n} = - \frac{1-t^{-n}}{n} \left[ \frac{1}{1-q^n} -
    \mathrm{Ch}_{\lambda} \left( q^n, \left( t/q \right)^n
    \right) + \left( \frac{t}{q} \right)^n \mathrm{Ch}_{\mu}
    \left( q^n, \left( t/q \right)^n \right) \right], \qquad \text{for
  } n \geq 1.}
\end{equation}

Having $c_n$ we have completely determined the vertex operator part
$\mathbf{R}^{\lambda}_{\mu}(w)$ of
$\widetilde{\mathcal{R}}^{\lambda}_{\mu}(w,u)$. One can check that for
$\lambda = \mu = \varnothing$ it reproduces the results
of~\cite{Zenkevich:2020ufs}.

\subsection{The prefactor}
\label{sec:finding-prefactor}
To determine $\widetilde{\mathcal{R}}^{\lambda}_{\mu}(w,u)$ completely
it remains to find the prefactor $f^{\lambda}_{\mu}(w,u)$. To this end
we have to consider the intertwining relation~\eqref{eq:15} with $g$
being the currents $x^{+}(z)$.

The coproduct of $x^{+}(z)$ is given by
\begin{equation}
  \label{eq:20}
    \Delta(x^{+}(z)) = x^{+}(z) \otimes 1 + \psi^{-}\left(
    \gamma_{(1)}^{\frac{1}{2}}z \right) \otimes x^{+} \left(
    \gamma_{(1)} z \right).
\end{equation}
Plugging Eq.~\eqref{eq:20} into the intertwining
relation~\eqref{eq:15} and sandwiching with the states on the vertical
representation we get
\begin{multline}
  \label{eq:21}
  f^{\lambda+1_i}_{\mu}(w,u) \sum_{i=1}^{l(\lambda)+1}
  A^{+}_{\lambda,i} ( q, t/q) \delta \left( \frac{z}{q^{\lambda_i} (t/q)^{i-1}
      w} \right) \mathbf{R}^{\lambda+1_i}_{\mu}(w,u) +\\
  + \frac{1}{\sqrt{t}} \exp \left[\sum_{n \geq 1} \frac{1}{n} \left(
      \frac{z}{w} \right)^n \left( 1 - t^n + \kappa_n
      \mathrm{Ch}_{\lambda} \left( q^{-n}, (t/q)^{-n} \right) \right)
  \right] f^{\lambda}_{\mu}(w,u) \mathbf{R}^{\lambda}_{\mu}(w)
  x^{+}(z)|_{\mathcal{F}^{(1,0)}_{q,t^{-1}}(u)} \stackrel{?}{=}\\
  \stackrel{?}{=} x^{+}(z)|_{\mathcal{F}^{(1,0)}_{q,t^{-1}}(u)}
  \mathbf{R}^{\lambda}_{\mu}(w) + f^{\lambda}_{\mu - 1_i}(w,u) \sum_{i=1}^{l(\mu)} \sqrt{t} q
  A^{-}_{\mu,i }( q, t/q) \delta \left( \frac{z}{q^{\mu_i -1} \left(
        \frac{t}{q} \right)^{i-1} w} \right) \left.\psi^{-}\left( \left(
      \frac{t}{q} \right)^{\frac{1}{4}} z
  \right)\right|_{\mathcal{F}^{(1,0)}_{q,t^{-1}}}
\mathbf{R}^{\lambda}_{\mu - 1_i} (w),
\end{multline}
where we have again used the formulas for the vertical DIM
representation from Appendix~\ref{sec:vert-repr-dim}, where one can
also find the definitions of $A^{\pm}_{\lambda,i}(q,t/q)$. 

Normal ordering the operators in every term of Eq.~\eqref{eq:21} we
notice that all the terms are proportional to
$:x^{+}(z)|_{\mathcal{F}^{(1,0)}_{q,t^{-1}}(u)}
\mathbf{R}^{\lambda}_{\mu}(w,u):$ while the coefficients are
\begin{multline}
  \label{eq:22}
  f^{\lambda+1_i}_{\mu}(w,u) \sum_{i=1}^{l(\lambda)+1}
  A^{+}_{\lambda,i} ( q, t/q) \delta \left( \frac{z}{q^{\lambda_i}
      (t/q)^{i-1}   w} \right) + \frac{u}{\sqrt{t}} f^{\lambda}_{\mu}(w,u) h^{\lambda}_{\mu}
  \left( \frac{z}{w} ; q^{-1}, t^{-1}\right) \stackrel{?}{=} \\
  \stackrel{?}{=} f^{\lambda}_{\mu - 1_i}(w,u) \sum_{i=1}^{l(\mu)} \sqrt{t} q
  A^{-}_{\mu,i }( q, t/q) \delta \left( \frac{z}{q^{\mu_i -1} \left(
        \frac{t}{q} \right)^{i-1} w} \right) + \sqrt{t} u
  f^{\lambda}_{\mu}(w,u) h^{\lambda}_{\mu}
  \left( \frac{w}{z} ; q, t\right),
\end{multline}
where we have introduced
\begin{multline}
  \label{eq:23}
  (1-q^{-1})(1-t^{-1}) h^{\lambda}_{\mu}(x;q,t) =\\
  = \exp \left[ \sum_{n \geq 1} \frac{1}{n}
    \left( \frac{w}{z} \right)^n (1-t^{-n}) \left( 1 + (1-q^n) \left\{
        - \mathrm{Ch}_{\lambda} \left( q^n, (t/q)^n \right) + \left(
          \frac{t}{q} \right)^n \mathrm{Ch}_{\mu} \left( q^n, (t/q)^n
        \right) \right\} \right) \right]=\\
  =\frac{1- \frac{1}{t} \frac{w}{z}}{1 - \frac{w}{z}} \prod_{(i,j) \in
  \lambda} \phi_{q,t/q} \left( x q^{j-1} (t/q) ^{i-1}\right) \prod_{(k,l) \in
  \mu} \frac{1}{\phi_{q^{-1},q/t} \left( x q^{l-1} (t/q) ^{k-1}\right)},
\end{multline}
and where
\begin{equation}
  \label{eq:26}
  \phi_{q,t/q}(x) = \frac{(1-x)\left( 1 - \frac{q}{t} x \right)}{(1-
    qx) \left( 1 - \frac{x}{t} \right)}.
\end{equation}
The equality~\eqref{eq:22} should rely on the following formula
involving the delta-function:
\begin{equation}
  \label{eq:25}
  h^{\lambda}_{\mu}(x;q,t) = \frac{1}{t}
  h^{\lambda}_{\mu}(x^{-1};q^{-1},t^{-1}) + \sum_{x_{*}\in
    \text{poles of } h^{\lambda}_{\mu}(x)} \delta \left( \frac{x}{x_{*}} \right)
  \mathrm{Res}_{x = x^{*}} \left( \frac{h^{\lambda}_{\mu}(x;q,t)}{x} \right) 
\end{equation}
To check Eq.~\eqref{eq:22} we need to ensure two conditions:
\begin{enumerate}
\item that poles of $h^{\lambda}_{\mu}\left(\frac{w}{z};q,t\right)$
  match the positions of the delta-functions in the two sums,
\item that the residues of
  $h^{\lambda}_{\mu}\left(\frac{w}{z};q,t\right)$ at the poles match
  the coefficients of the corresponding delta-functions.
\end{enumerate}

The dependence of $f^{\lambda}_{\mu}(w,u)$ on $u$ and $w$ can be
determined immediately from Eq.~\eqref{eq:22}. Indeed, the residues
like those in~\eqref{eq:25} don't depend on $w$ and $u$ enters
in~\eqref{eq:22} only in the second terms in both lines. Thus
\begin{equation}
  \label{eq:24}
\boxed{  f^{\lambda}_{\mu} (w,u) = u^{|\lambda| - |\mu|} f^{\lambda}_{\mu},}
\end{equation}
where $f^{\lambda}_{\mu}$ in the r.h.s.\ doesn't depend on the
spectral parameters (though does still implicitly depend on $q$ and
$t$).

After experimenting a little bit with the pole structure of
$h^{\lambda}_{\mu}(w/z;q,t)$ one learns that for certain pairs of Young
diagrams $(\lambda,\mu)$ there appear to be spurious poles which never
appear in the delta-functions. It turns out that these spurious poles
appear precisely when partition $\lambda$ does not \emph{interlace}
partition $\mu$. We say that $\lambda$ interlaces $\mu$ and write
$\lambda \succ \mu$ if
\begin{equation}
  \label{eq:27}
  \ldots \geq \lambda_i \geq \mu_i \geq \lambda_{i+1} \geq \mu_{i+1}
  \geq \ldots
\end{equation}
Notice that $\lambda \succ \mu$ implies $\lambda \supseteq \mu$, but
the former condition is much more restrictive.

The only way to get rid of the spurious poles is to set
\begin{equation}
  \label{eq:28}
\boxed{  f^{\lambda}_{\mu} = 0 \quad \text{if } \lambda \nsucc \mu,}
\end{equation}
so that the whole term $f^{\lambda}_{\mu}h^{\lambda}_{\mu}(w/z;q,t)$
vanishes. Then we need to check the consistency of~\eqref{eq:28} by
looking at the case when $\lambda \succ \mu$, but for some $i$ we get
$\lambda + 1_i \nsucc \mu$ or $\lambda \nsucc \mu - 1_i$. This would
mean that the coefficients in front of the corresponding
delta-functions vanish, but
$f^{\lambda}_{\mu}h^{\lambda}_{\mu}(w/z;q,t)$ (in which
$f^{\lambda}_{\mu} \neq 0$) still potentially contains such poles. It
turns out, however, that the pole structure of
$h^{\lambda}_{\mu}(w/z;q,t)$ guarantees that these poles do not
arise. Indeed, one can check the following explicit expression for
$h^{\lambda}_{\mu}(w/z;q,t)$ which is valid for $\lambda \succ \mu$:
\begin{equation}
  \label{eq:29}
  h^{\lambda}_{\mu}(w/z;q,t) = \prod_{
    \begin{smallmatrix}
      (i,\lambda_i) \in
    \text{addable}(\lambda)\\
\lambda+1_i \succ \mu
    \end{smallmatrix}} \frac{1 - \frac{1}{t} \frac{w}{z} q^{\lambda_i}
  \left(\frac{t}{q}\right)^{i-1}}{1 -  \frac{w}{z} q^{\lambda_i}
  \left(\frac{t}{q}\right)^{i-1}} \prod_{
    \begin{smallmatrix}
      (k,\mu_k) \in
    \text{removable}(\mu)\\
\lambda \succ \mu-1_i
    \end{smallmatrix}} \frac{1 - t \frac{w}{z} q^{\mu_i-1}
  \left(\frac{t}{q}\right)^{i-1}}{1 -  \frac{w}{z} q^{\mu_i-1}
  \left(\frac{t}{q}\right)^{i-1}}.
\end{equation}
Thus the poles of $h^{\lambda}_{\mu}(w/z;q,t)$ corresponding to
$\lambda + 1_i \nsucc \mu$ or $\lambda \nsucc \mu - 1_i$ vanish.

The determination of $f^{\lambda}_{\mu}$ reduces to the solution of
the following recurrence relations:
\begin{align}
  \label{eq:30}
  f^{\lambda}_{\mu} \mathrm{Res}_{z=w q^{\lambda_i} \left( \frac{t}{q}
    \right)^{i-1}} \frac{1}{z} h^{\lambda}_{\mu} \left( \frac{w}{z};
    q,t \right) &\stackrel{?}{=} f^{\lambda + 1_i}_{\mu}
  A^{+}_{\lambda,i} \left( q, \frac{t}{q} \right),\\
  f^{\lambda}_{\mu} \mathrm{Res}_{z=w q^{\mu_i-1} \left( \frac{t}{q}
    \right)^{i-1}} \frac{1}{z} h^{\lambda}_{\mu} \left( \frac{w}{z};
    q,t \right) &\stackrel{?}{=} f^{\lambda}_{\mu- 1_i}
  q A^{-}_{\mu,i} \left( q, \frac{t}{q} \right).
\end{align}
We have checked the consistency of the
relations~\eqref{eq:30},~\eqref{eq:24} and solved them for lower Young
diagrams, for example
\begin{equation}
  \label{eq:43}
  f^{[k]}_{\varnothing} = 1.
\end{equation}
It would be desirable to obtain a general solution for
$f^{\lambda}_{\mu}$.

\section{Statistical model and plane partitions}
\label{sec:comp-plane-part}
We can combine the two types of crossing operators considered in
sec.~\ref{sec:warm-up:-degenerate},~\ref{sec:hanany-witten-5} (the
degenerate resolution and the Hanany-Witten crossing respectively)
into a version of a lattice statistical model. In such a model the
states are Young diagrams and live on the edges of the square lattice,
while the crossings play the roles of Boltzmann weights at the
quadrivalent vertices. The model includes additional data: the every
line of the lattice carries a spectral parameter and is colored either
violet, red or blue. Crossings of all possible types are obtained from
the results of
sec.~\ref{sec:warm-up:-degenerate},~\ref{sec:hanany-witten-5} by
permuting the parameters $q$, $t^{-1}$ and $t/q$.

The general theory of such statistical models will be investigated
elsewhere\footnote{Similar models have appeared
  in~\cite{FJMM-Bethe-2}.}. Here we consider the simplest element of
the lattice model --- the transfer matrix, i.e.\ the combination of
$N$ crossings, which we contract vertically along a single line. We
consider two cases:
\begin{enumerate}
\item When the vertical and horizontal branes are of the same color:
  \begin{equation}
\mathcal{T}^{\lambda^{(1)}}_{\varnothing} (\vec{u},w) \quad = \quad   \includegraphics[valign=c]{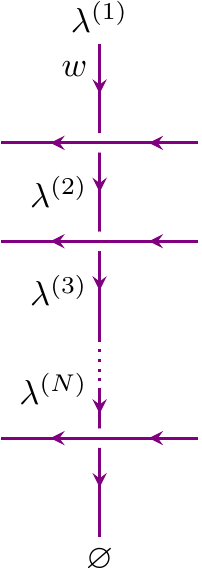}\label{eq:44}
\end{equation}
Because of the selection rules discussed in
sec.~\ref{sec:warm-up:-degenerate} the partitions $\lambda^{(i)}$ on
the intermediate edges giving nontrivial contribution to the transfer
matrix satisfy
\begin{equation}
  \label{eq:31}
  \lambda^{(1)} \supseteq \lambda^{(2)} \supseteq \ldots \supseteq
  \lambda^{n} \supseteq \varnothing.
\end{equation}
Therefore, the ordered set $(\lambda^{(1)}, \lambda^{(2)}, \ldots
\lambda^{(n)})$ can be viewed collectively as a $3d$ Young diagram or
plane partition $\pi$ of height not more than $N$ and fixed first
slice. The transfer matrix is an operator acting on
$\bigotimes_{i=1}^n\mathcal{F}^{(1,0)}_{q,t^{-1}}(u_i)$ then given by
the sum over such plane partitions:
\begin{multline}
  \label{eq:46}
  \mathcal{T}^{\lambda^{(1)}}_{\varnothing} (\vec{u},w) = \sum_{
    \begin{smallmatrix}
      \pi\\
      \mathrm{height}(\pi) \leq N\\
      \pi^{(1)} = \lambda^{(1)}
    \end{smallmatrix}
}
  C_{\pi} \prod_{i=1}^N u_i^{|\pi^{(i)}|-|\pi^{(i+1)}|} : \exp \left[ - \sum_{n \geq 1} \frac{w^n}{n} \frac{\left( 1-
        \left( t/q \right)^n\right)}{(1-q^n)} \sum_{i=1}^N
    (t/q)^{\frac{n(i-1)}{2}} a_{-n}^{(i)}\right] \times \\
  \times \exp \left[ - \sum_{n \neq 0} \frac{(1-t^n) }{n} 
      \mathrm{Ch}_{\pi^{(1)}}(q^{-n},t^n) w^{-n} a_n\right]\times\\
  \times \exp
  \left[ - \sum_{n \neq 0} \frac{(1-t^n) }{n}
    \sum_{i=1}^N(t/q)^{\frac{n(1-i)}{2}} \mathrm{Ch}_{\pi^{(i+1)}}(q^{-n},t^n) \left(
       a^{(i+1)} - \left( \frac{t}{q}
      \right)^{\frac{|n|}{2}} a_n^{(i)} 
    \right)w^{-n}  \right] :,
\end{multline}
where $C_{\pi}$ are certain explicit coefficients and $a^{(i)}_n$ are
modes acting on the $i$-the horizontal Fock space.

\item When the vertical and horizontal branes have different colors.
  \begin{equation}
    \label{eq:45}
\widetilde{\mathcal{T}}^{\lambda^{(1)}}_{\varnothing} (\vec{u},w) \quad = \quad    \includegraphics[valign=c]{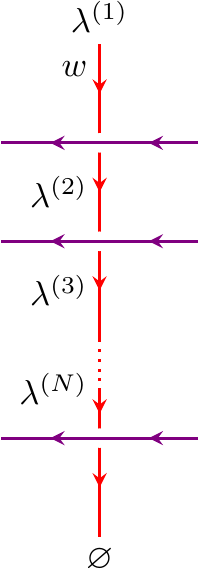}
  \end{equation}
  Then, according to sec.~\ref{sec:hanany-witten-5} the diagrams
  $\lambda^{(i)}$ satisfy the interlacing condition:
  \begin{equation}
    \label{eq:47}
    \lambda^{(1)} \succ \lambda^{(2)} \succ \ldots \succ
  \lambda^{n} \succ \varnothing.
\end{equation}
In fact, the conditions~\eqref{eq:47} also allow us to view the set
$(\lambda^{(1)}, \lambda^{(2)}, \ldots \lambda^{(n)})$ as (a part of)
a plane partition. However, this time $\lambda^{(i)}$ correspond to
\emph{diagonal} slicing, i.e.
\begin{equation}
  \label{eq:48}
  \lambda^{(i)}_j = \pi^{(j)}_{i+j-1}.
\end{equation}
One can check that the interlacing conditions~\eqref{eq:47} are
equivalent to the definition of plane partition.

The resulting transfer-matrix is given by the sum over ``halves of
plane partitions'', i.e.\ we slice a plane partition along the plane
$x=y$ in $3d$ space, remove one half of it and fix the boundary
condition on the slice to be $\lambda^{(1)}$. The resulting partition
function is
\begin{multline}
  \label{eq:49}
 \widetilde{\mathcal{T}}^{\lambda^{(1)}}_{\varnothing} (\vec{u},w)= \sum_{
    \begin{smallmatrix}
      \pi\\
      \mathrm{height}(\pi) \leq N\\
      \pi^{(i)}_i = \lambda^{(1)}_i
    \end{smallmatrix}
  }\tilde{C}_{\pi} \prod_{i=1}^N u_i^{|\pi^{(i)}|-|\pi^{(i+1)}|} :
  \exp \left[ - \sum_{n \geq 1} \frac{w^n}{n} \frac{\left( 1-
        t^{-n}\right)}{(1-q^n)} \sum_{i=1}^N
    (t/q)^{\frac{n(i-1)}{2}} a_{-n}^{(i)}\right] \times \\
  \times \exp \left[ - \sum_{n \neq 0} \frac{(1-t^n) }{n}
    \mathrm{Ch}_{\pi^{(1)}}(q^{-n},(t/q)^{-n}) w^{-n} a_n^{(1)}\right]\times\\
  \times \exp \left[ - \sum_{n \neq 0} \frac{(1-t^n) }{n}
    \sum_{i=1}^N(t/q)^{\frac{n(1-i)}{2}}
    \mathrm{Ch}_{\pi^{(i+1)}}(q^{-n},(t/q)^{-n}) \left( a^{(i+1)}_n - \left(
        \frac{t}{q} \right)^{\frac{|n|}{2}} a_n^{(i)} \right)w^{-n}
  \right] :
\end{multline}
where $\tilde{C}_{\pi}$ are again certain coefficients.

\end{enumerate}

Another potential building block of a lattice model is the
\emph{trace} of the transfer matrix, i.e.\ the \emph{compactification}
of the network of intertwiners like~\eqref{eq:44}
or~\eqref{eq:45}. Due to the selection rules~\eqref{eq:31}
and~\eqref{eq:47} if we set $\lambda^{(1)} = \lambda^{(N)}$ to enforce
the compactification we find that all the intermediate Young diagrams
become equal, $\lambda^{(i)} = \lambda^{(1)}$. Thus, the trace is
given by the sum over single Young diagrams, not plane partitions.

However, there is one more way to compactify the
diagrams~\eqref{eq:44} or~\eqref{eq:45} which might seem strange at
first but turns out to be very natural. The idea is to identify all
horizontal lines and wrap the vertical line many times around the
resulting cylinder as a spiral:
\begin{equation}
  \label{eq:50}
      \includegraphics[valign=c]{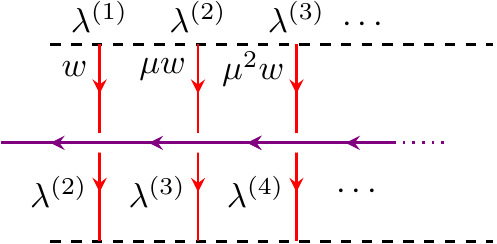}
\end{equation}
Such operator is a sum over halves of plane partitions but it acts on
a single Fock space. It can be obtained from Eq.~\eqref{eq:49} by
the identification
\begin{align}
  \label{eq:51}
  a^{(i)}_n &\mapsto \mu^{i-1} a_n,\\
  u_i &\mapsto \Lambda^{i-1} u.
\end{align}
We get
\begin{multline}
  \label{eq:53}
  \widetilde{\mathcal{R}}_{\mathrm{spiral}}^{\lambda^{(1)}} = \sum_{
    \begin{smallmatrix}
      \pi\\
      \pi^{(i)}_i = \lambda^{(1)}_i
    \end{smallmatrix}
  }\tilde{C}_{\pi} \prod_{i=1}^N \Lambda^{|\pi|} :
  \exp \left[ - \sum_{n \geq 1} \frac{w^n}{n} \frac{\left( 1-
        t^{-n}\right)}{(1-q^n) \left( 1 - \left(\mu \sqrt{\frac{t}{q}}\right)^n \right)}  a_{-n}\right] \times \\
  \times \exp \left[ - \sum_{n \neq 0} \frac{(1-t^n) }{n}
    \mathrm{Ch}_{\pi^{(1)}}(q^{-n},(t/q)^{-n}) w^{-n} a_n\right]\times\\
  \times \exp \left[ - \sum_{n \neq 0} \frac{(1-t^n) \left( 1- \left(
        t/q \right)^{\frac{|n|}{2}} \mu^{-n} \right)}{n}
    \sum_{i=1}^N(t/q)^{\frac{n(1-i)}{2}}
    \mathrm{Ch}_{\pi^{(i+1)}}\left(q^{-n},\left(\frac{q}{t}\right)^n, \left( \mu \sqrt{\frac{q}{t}} \right)^n \right)   a_n w^{-n}
  \right] :
\end{multline}
where the generalization of Eq.~\eqref{eq:33} to plan partitions reads
\begin{equation}
  \label{eq:54}
  \mathrm{Ch}_{\pi}(q_1,q_2,q_3) = \sum_{(i,j,k)\in \pi} q_1^{k-1}
  q_2^{i-1} q_3^{k-1}.
\end{equation}

Algebraically the operator~\eqref{eq:50} is an infinite product of
$R$-matrices with shift operators $\mu^d \Lambda^{d_{\perp}}$
sandwiched in between:
\begin{equation}
  \label{eq:52}
  \widetilde{\mathcal{R}}_{\mathrm{spiral}} \sim \prod_{i\geq 1} \mathcal{R}\times \mu^d \Lambda^{d_{\perp}}.
\end{equation}
Such $R$-matrices have appeared in the study of elliptic deformations
of quantum affine Lie algebras~\cite{Jimbo:1999zz}\footnote{For
  another connection between compactification and elliptic deformation
  of the DIM algebra see~\cite{Ghoneim:2020sqi}}.

We will consider the spiral setup in more detail in~\cite{future}
where we will show that it is related to $qq$-characters of affine
root system of type $\widehat{A}_0$.

\section{Conclusions}
\label{sec:conclusions}
We have presented an explicit expression for a new type of
intertwining operator of DIM representations, corresponding to the
Hanany-Witten type brane crossing. The operator is in fact the
universal DIM $R$-matrix evaluated in the Fock representations
corresponding to the crossing branes. It has interesting combinatorial
properties involving interlacing partitions and can be used to build
statistical models corresponding to systems of 5-brane which
generalize 5-brane webs. We have built transfer matrices of these
lattice models and have identified certain ``spiral'' compactification
of the transfer matrices which are given by sums of vertex operators
labelled by plane partitions.

The field theory meaning of the crossing operator is not completely
clear yet, but should naturally involve gauge theories living on two
different $\mathbb{C}^2$ spaces interacting along a shared
$\mathbb{C}$ plane.

\section*{Acknowledgements}
\label{sec:acknowledgements}
This work is partly supported by the joint grant RFBR 21-51-46010 and
T\"{U}BITAK 220N106.

\appendix

\section{Vertical representation of DIM}
\label{sec:vert-repr-dim}
For completeness we present here the action of the DIM algebra
currents in the vertical Fock representation
$\mathcal{F}^{(0,1)}_{q,t/q}(w)$. We have
\begin{align}
  \label{eq:32}
  x^{+}(z)|_{\mathcal{F}^{(0,1)}_{q,t/q}(w)} |\lambda,w\rangle &=
  \sum_{i=1}^{l(\lambda)+1} A^{+}_{\lambda,i} \left( q, \frac{t}{q}
  \right) \delta \left( \frac{z}{w q^{\lambda_i} (t/q)^{i-1}} \right)
  |\lambda+1_i,w \rangle,\\
  x^{-}(z)|_{\mathcal{F}^{(0,1)}_{q,t/q}(w)} |\lambda,w\rangle &=
  \sum_{i=1}^{l(\lambda)} A^{-}_{\lambda,i} \left( q, \frac{t}{q}
  \right) \delta \left( \frac{z}{w q^{\lambda_i-1} (t/q)^{i-1}} \right)
  |\lambda-1_i,w \rangle, \label{eq:34}\\
  \psi^{+}(z)|_{\mathcal{F}^{(0,1)}_{q,t/q}(w)} |\lambda,w\rangle &=
  \sqrt{t}\exp \left[ \sum_{n \geq 1} \frac{1}{n} \left(
      \frac{w}{z} \right)^n \left( 1 - t^{-n} - \kappa_n
      \mathrm{Ch}_{\lambda} (q^n, (t/q)^n) \right) \right],\\
  \psi^{-}(z)|_{\mathcal{F}^{(0,1)}_{q,t/q}(w)} |\lambda,w\rangle &=
  \frac{1}{\sqrt{t}}\exp \left[ \sum_{n \geq 1} \frac{1}{n} \left(
      \frac{z}{w} \right)^n \left( 1 - t^n + \kappa_n
      \mathrm{Ch}_{\lambda} (q^{-n}, (q/t)^n) \right) \right]
\end{align}
where $\kappa_n = (1-q^n)(1-t^{-n})(1-(t/q)^n)$,
$\mathrm{Ch}_{\lambda}(q,t)$ is defined in Eq.~\eqref{eq:33} and
\begin{align}
  \label{eq:35}
  A^{+}_{\lambda,i}\left( q, \frac{t}{q} \right) &= \frac{1}{1-q^{-1}} \prod_{j=1}^i \psi \left(
    q^{\lambda_i - \lambda_j} (t/q)^{i-j} \right),\\
  A^{-}_{\lambda,i}\left( q, \frac{t}{q} \right) &= - \frac{t^{-\frac{1}{2}}}{1-q} \frac{1 - 
    q^{\lambda_i} }{1 - \frac{1}{t}
    q^{\lambda_i} } \prod_{j=i+1}^{l(\lambda)} \frac{\psi \left(
    q^{\lambda_i - \lambda_j-1} \left(\frac{t}{q}\right)^{i-j} \right)}{\psi \left(
    q^{\lambda_i - 1} \left(\frac{t}{q}\right)^{i-j} \right)}
\end{align}
where
\begin{equation}
  \label{eq:36}
  \psi(x) = \frac{(1 - t x) \left( 1 - \frac{q}{t} x\right)}{(1-x) (1
    - q x)}.
\end{equation}

The notation $\lambda \pm 1_i$ in Eqs.~\eqref{eq:32},
and~\eqref{eq:34} means a diagram obtained from $\lambda$ by adding a
box at $(\lambda_i, i)$. The coefficients $A^{\pm}_{\lambda,i}$
conveniently vanish when the resulting diagram is not a Young diagram,
so we don't have to introduce any special conditions on where the box
is added.

The transition coefficients $A^{\pm}_{\lambda,i}$ satisfy the
following identity:
\begin{equation}
  \label{eq:37}
  A^{+}_{\lambda,i} \left( q, \frac{t}{q} \right) = \sqrt{t} q\,
  A^{-}_{\lambda+1_i,i} \left( q, \frac{t}{q} \right)
  \frac{C'_{\lambda}\left( q, \frac{t}{q} \right) C_{\lambda + 1_i}
    \left( q, \frac{t}{q} \right)}{C_{\lambda}\left( q, \frac{t}{q}
    \right) C'_{\lambda+1_i}\left( q, \frac{t}{q} \right)},
\end{equation}
where
\begin{align}
  \label{eq:38}
  C_{\lambda}\left(q,\frac{t}{q}\right) &= \prod_{(i,j)\in \lambda} \left( 1 - q^{\lambda_i - j}
    \left( \frac{q}{t} \right)^{\lambda_j^{\mathrm{T}} - i +1}
  \right),\\
    C'_{\lambda}\left(q,\frac{t}{q}\right) &= \prod_{(i,j)\in \lambda} \left( 1 - q^{\lambda_i - j+1}
    \left( \frac{q}{t} \right)^{\lambda_j^{\mathrm{T}} - i} \right).
\end{align}

Eq.~\eqref{eq:37} implies that if we introduce the standard Macdonald
scalar product on $\mathcal{F}^{(0,1)}_{q,t/q}(w)$,
\begin{equation}
  \label{eq:39}
  \langle \lambda, w | \mu, w\rangle =  \delta_{\lambda,\mu} \frac{C'_{\lambda}\left(q,\frac{t}{q}\right)}{C_{\lambda}\left(q,\frac{t}{q}\right)},
\end{equation}
then the currents $\psi^{\pm}(z)$ are self-adjoint and $x^{\pm}(z)$
are adjoints of each other:
\begin{align}
  \label{eq:40}
  \left( \psi^{\pm}(z) \right)^{\dag} &= \psi^{\pm}(z),\\
  \left( x^{+}(z) \right)^{\dag} &= \sqrt{t} q\, x^{-}(z).
\end{align}

There is another convenient formula expressing the coefficients
$A^{\pm}_{\lambda,i}$ as residues:
\begin{align}
  \label{eq:41}
  A^{+}_{\lambda,i}\left( q, \frac{t}{q} \right) &= \frac{\left(
      \frac{q}{t} \right)^{\lambda_i+1}}{1-q^{-1}}
  \frac{C'_{\lambda+1_i}\left( q, \frac{t}{q}
    \right)}{C'_{\lambda}\left( q, \frac{t}{q} \right)} \mathrm{Res}_{z
    = q^{\lambda_i} (t/q)^{i-1}} \left\{ \frac{1}{z(1-z)}
    \prod_{(k,l)\in \lambda} \phi \left( \frac{z}{q^{l-1}(t/q)^{k-1}}
    \right)\right\},\\
  A^{-}_{\lambda,i}\left( q, \frac{t}{q} \right) &=
  \frac{\sqrt{t} \left(
      \frac{t}{q} \right)^{\lambda_i}}{1-t} \frac{C'_{\lambda-1_i}\left( q, \frac{t}{q}
    \right)}{C'_{\lambda}\left( q, \frac{t}{q} \right)} \mathrm{Res}_{z
    = q^{\lambda_i-1} (t/q)^{i-1}} \left\{ \frac{(1-z)}{z}
    \prod_{(k,l)\in \lambda} \frac{1}{\phi \left( \frac{z}{q^{l-1}(t/q)^{k-1}}
    \right)}\right\}  
\end{align}
where
\begin{equation}
  \label{eq:42}
  \phi(x) = \frac{(1-x) \left( 1 - \frac{1}{t} x\right)}{\left( 1 -
      \frac{x}{q} \right) (1 - \frac{q}{t} x)}.
\end{equation}

\end{document}